\documentclass[aps,prd,eqsecnum,nofootinbib,twocolumn]{revtex4}

\usepackage{amsmath}
\usepackage{graphicx,epsfig}
\usepackage{mathrsfs}
\usepackage{amssymb}
\usepackage{anysize}
\usepackage{slashed}
\usepackage{epsfig}
\usepackage{color}
\usepackage[dvipsnames]{xcolor}

\newcommand{\be}{\begin{equation}}
\newcommand{\ee}{\end{equation}}
\newcommand{\bea}{\begin{eqnarray}}
\newcommand{\beas}{\begin{eqnarray*}}
\newcommand{\eea}{\end{eqnarray}}
\newcommand{\eeas}{\end{eqnarray*}}
\newcommand{\ba}{\begin{array}}
\newcommand{\ea}{\end{array}}
\newcommand{\Mt}{\widetilde{M}}
\newcommand{\mt}{\widetilde{m}}
\newcommand{\mA}{\Delta m_{ATM}^{2}}
\newcommand{\mS}{\Delta m_{\odot}^{2}}

\def\ls{\mathrel{\lower4pt\vbox{\lineskip=0pt\baselineskip=0pt
           \hbox{$<$}\hbox{$\sim$}}}}
\def\gs{\mathrel{\lower4pt\vbox{\lineskip=0pt\baselineskip=0pt
           \hbox{$>$}\hbox{$\sim$}}}}
\def\la{\langle}
\def\ra{\rangle}
\begin{document}

\title{Dirac neutrino mixings from hidden $\mu - \tau$ symmetry}

\author{Edgar R. Luna Terrazas\footnote{email: eluna@fis.cinvestav.mx},
 Abdel P\'erez-Lorenzana\footnote{email: aplorenz@fis.cinvestav.mx}}

\affiliation{
$^1$Departamento de F\'{\i}sica, Centro de Investigaci\'on y de Estudios
Avanzados del I.P.N.\\
Apdo. Post. 14-740, 07000, Ciudad de M\'exico, M\'exico}

\date{\today}

\date{\today}

\begin{abstract}
We explore masses and mixings for Dirac neutrinos in models where lepton 
number is conserved, under the guidance of a hidden, but broken,  $\mu-\tau$ 
exchange symmetry,  that makes itself evident in the squared hermitian mass 
matrix. We study the parameter space in the most general theory as allowed by  
current neutrino oscillation experiment data. 
By using a general parameterization of the mass matrix which contains only 
observable parameters 
we stablish that the amount of breaking of the symmetry is 
in the range of the atmospheric mass scale, without regard to the neutrino 
hierarchy, the absolute neutrino mass and the Dirac CP phase. 
An estimate of the invisible branching ratio for a Higgs boson 
decaying into Dirac neutrinos, $H\rightarrow\nu\overline{\nu}$,  is 
given and compared to recent measurements in this context.
\end{abstract}

%\keywords{}
%\packs{ }
%\vskip90pt
\maketitle
%%%%%%%%%%%%%%%%%%%%%%%%%%%%%%%%%%%%%%%%%%%%%%%%%%%%%%%%%%%%%%%%%%%%%%
\section{Introduction}
%%%%%%%%%%%%%%%%%%%%%%%%%%%%%%%%%%%%%%%%%%%%%%%%%%%%%%%%%%%%%%%%%

Neutrino oscillations have been for some time under the scope of a large number 
of theoretical studies and many experimental efforts, since they imply that 
neutrinos have mass and leptons mix flavors (for a review see for 
instance~\cite{PDG}). As the Standard Model (SM), on the 
contrary,  predicts that neutrinos should be rather  massless and no 
flavor mixing should exist in the lepton sector, neutrino physics seem to 
point towards the need of new physics. 
Compelling evidence for neutrino  
oscillations has been provided by data obtained from the observation of 
neutrinos arriving from the sun, from upper atmosphere interactions of cosmic 
rays, from nuclear reactors, and from particle accelerators. Most of such data  
can be understood in a framework with three weak flavor neutrinos, $\nu_\ell$ 
for $\ell =e,\mu,\tau$. Those corresponding to the three SM charged leptons. 

Oscillation phenomena is not sensible to the actual mass 
of the neutrinos but to their squared mass differences, 
$\Delta m^2_{ab} = m_a^2-m_b^2$, for $a,b = 1,2,3$. It is so, however, 
to the  mixing that connects the weak to the mass eigenstates, $\nu_a$. 
In the two neutrino flavor approximation, the 
oscillation probability is simply given as 
$P_{\ell\ell'} = \sin^2{2\theta}\sin^2(\Delta m_{ab}^2 E/4L)$,  for the 
mixing angle $\theta$. 
%This 
%approximation comes out to be accurate even for three flavors due to 
%the observed hierarchy in the relevant mass scales.  
Global fits~\cite{PDG,nuglobal}, with all three 
neutrinos, find for solar neutrino oscilations the scale
$\Delta m_{\odot}^2=\Delta m_{21}^2= m_2^2 -m_1^2 \sim 7.5 \times 10^{-5} 
eV^2$, 
whereas for atmospheric neutrinos they give
$\Delta m_{ATM}^2 = |\Delta m_{31}^2| \sim 2.5\times 
10^{-3} eV^2$. Note that the hierarchy among the first two mass eigenstates is 
well known due to the contribution of matter effects on solar 
neutrino oscillations. However,
the sign in $\Delta m_{31}^2= m_3^2 -m_1^2$, and therefore the neutrino mass 
hierarchy pattern, is still unknown. 
Data is so far  consistent with both 
normal 
($\Delta m_{31}^2 >0$) and inverted ($\Delta m_{31}^2 <0$) hierarchies.
As for the mixing angles, global fits indicate that 
$ \sin^2\theta_\odot\approx 0.308\pm 0.017$ for solar, 
$ \sin^2\theta_{ATM} \approx 
0.437^{+0.033}_{-0.023}~ 
(0.455^{+0.039}_{-0.031})$ for atmospheric, and 
$\sin^2\theta_{13}\approx 
0.0234^{+0.0020}_{-0.0019} ~ (0.0240^{+0.0019}_{-0.0022})$ for reactor 
neutrino oscillations with normal (inverted) hierarchy.

Despite the incontrovertible evidence for neutrino 
masses and mixings, none can be said still about the actual nature of the 
neutrino. The question regards to whether neutrino is its own 
antiparticle, in which case it is called a Majorana particle,  or not, in which 
case it would be a Dirac particle. 
Neutrino oscillations are consistent with both the possibilities.
Although Majorana neutrinos have an ideal signature 
on neutrinoless double beta decay, the experimental 
evidece for such processes is still lacking (for a review see for 
instance Ref.~\cite{ndbeta}). 
On the theoretical side, Majorana neutrinos are considered to be 
easier to understand, as  the seesaw mechanism~\cite{seesaw} can generate 
very small masses for the standard left handed neutrinos, at the account of 
introducing  large masses for right handed ones, without assuming any small 
value for the Yukawa couplings to the Higgs field. Nevertheless, appart from 
this naturalness argument, there is no other theoretical or experimental reason 
to believe that the other possibility can be ruled out, and so it remains. 
Yet, Dirac neutrino nature could be understood if total lepton number were a 
conserved quantity and some models oriented to account for the smallness of the 
neutrino mass in this case had already been explored (see for 
instance~\cite{dnmodels}).

On the other hand, there is the intriguing   
observation, of particular interest for model building,  that the measured 
mixing angles  comply with the empyrical relation
\be
1/2 - \sin^2\theta_{ATM}\approx \sin\theta_{13}/\text{few}~, 
\label{relation}
\ee
which suggests that the deviation of $\theta_{ATM}$ 
from its maximal value, $\Delta\theta=\pi/4 -|\theta_{ATM}|$,  could 
be correlated to the non zero value of $\theta_{13}$.
This can  be thought to indicate a possible common physical origin for both 
angles, since null values of  $\Delta\theta$ and $\theta_{13}$ do 
increase the symmetry in the Majorana neutrino mass sector
by exhibiting a discrete $\mu- \tau$ exchange symmetry~\cite{mutau1,mutau}. 
This has suggested the idea that observed values could be understood as a 
result of the breaking of $\mu-\tau$ symmetry. 
Many theoretical studies had been inspired by this observation in the last 
years~\cite{mutau,others,cpmutau,mtreview}, but few attention has been given 
to exploring the possibilities of this symmetry  for the description of Dirac 
neutrino mixings. Such is the main goal of  the present work. 

The outcome of our study has some interesting results
that we will discuss below. First of all, alike to what happens in 
the Majorana neutrino case, $\mu-\tau$ symmetry appears to be quite natural in 
the Dirac neutrino framework.  It of course arises from the condition of zero 
$\Delta\theta$ and $\theta_{13}$, which implies that the symmetry is rather 
broken. However, unlike the Majorana case (see for instance 
Ref.~\cite{cpmutau}), 
current experimental results 
indicate that the breaking of the symmetry is always relatively small for Dirac 
neutrinos, when compared to the heaviest neutrino mass, regardless of neutrino 
hierarchy and  the Dirac CP phase value.
Therefore, $\mu-\tau$ must be regarded as a good 
approximate symmetry.
This suggest that any realistic model built to provide Dirac neutrino masses 
and 
mixings  should contain $\mu-\tau$ symmetry as an implicit, or explicit, 
flavor symmetry.  

To state our case, we organize the present discussion as follows.
We start by revisiting the origin of Dirac masses in a lepton conserving 
extension of the SM, and introducing a  phenomenological 
parameterization for the Yukawa couplings that uses only all known experimental 
observables. We argue that those are the only physical parameters that are 
relevant to reconstruct the most general couplings in the lepton sector. 
Next we use this results to study $\mu-\tau$ symmetry 
and to parameterize its breaking in the context of Dirac neutrinos.
Furthermore, we use the above mentioned experimental results on neutrino 
masses and mixings to explore the breaking parameter space in order to 
stablish the amount of breaking allowed for the experimental data.
We also calculate explicit expressions for 
$\Delta\theta$ and $\theta_{13}$, in the limit of a small breaking of the 
symmetry to evidence their correlation under the $\mu-\tau$ symmetric 
approach. 
Some comments about the implications of our Yukawa couplings parameterization 
on the invisible width of the Higgs are also made. 
Some further discussion and our conclusions are finally 
presented.

%%%%%%%%%%%%%%%%%%%%%%%%%%%%%%%%%%%%%%%%%%%%%%%%%%%%%%%%%%%%%%%%%
\section{A phenomenological parameterization for Yukawa couplings}
%%%%%%%%%%%%%%%%%%%%%%%%%%%%%%%%%%%%%%%%%%%%%%%%%%%%%%%%%%%%%%%%%
Extending the SM with the simple addition of  right handed siglet neutrinos, 
$N_\ell$,
introduces an anomaly free global symmetry in the theory, which is associated 
to the combination of baryon and total lepton numbers, $B-L$. 
As Majorana mass terms violate the conservation of lepton number by two units, 
they are not possible if $L$ (that is, $B-L$) is 
assumed to be conserved. Under such an assumption, the most general Yukawa 
couplings are written as
\be
f_\ell \bar L_\ell H \ell_R + y_{\ell\ell^\prime} \bar L_\ell \widetilde{H} 
N_{\ell^\prime} + h.c.
\label{yukawas}
\ee
They would be responsible for lepton masses through the Higgs mechanism. Here 
$L$ stands for the standard left handed lepton doublets, $H$ for the Higgs, and 
$\ell_R$ for the standard right handed charged leptons.
Notice that, without loss of generality, we have chosen to work in the basis 
where charge lepton couplings are already diagonal. This can always be made.
If a specific model for lepton masses were to provide non diagonal  Yukawa 
couplings in that sector, one can always trasform it into the expression 
given above by picking up the specific  $U(3)_L\times U(3)_R$ 
flavor transformations, 
$L_\ell\rightarrow O^L_{\ell\ell^\prime}L_{\ell^\prime}$ and $\ell_R\rightarrow 
O^R_{\ell\ell^\prime}\ell_R^\prime$,  that diagonalize the corresponding 
Yukawa matrix, under which all other SM terms are invariant, but for the Yukawa 
couplings that involve the right handed neutrinos. These last would just be 
properly redefined by the $O^L$ transformation. We work on such a basis 
on what follows. After that electroweak symmetry breaking is introduced,
one gets the most general Dirac mass terms as
\be
m_\ell \bar\ell_L\ell_R + (M_\nu)_{\ell\ell^\prime} \bar\nu_\ell N_\ell^\prime 
+ h.c.~,
\ee
where, clearly,  $m_\ell = f_\ell v$ and $(M_\nu)_{\ell\ell^\prime} = 
v y_{\ell\ell^\prime}$,  with $v = \la H\ra$ the Higgs vacuum expectation 
value. Note that 
Dirac neutrino mass matrix, $M_\nu$, is non diagonal and complex in general. 
Its diagonalization is done through a bi-unitary transformation, that 
requires the simultaneous transformation of left and right handed neutrino 
flavor spaces. Thus, we take 
\be 
\nu_\ell = U_{\ell a} \nu_a \qquad \text{and}\qquad 
N_{\ell} = V_{\ell a} N_a~, \label{rotations} 
\ee
such that in the new basis  
the neutrino mass terms become
$ (M_d)_{ab} \bar\nu_a N_b +h.c.$, where 
\be
M_d = U^\dagger M_\nu V= diag\{m_1,m_2,m_3\}~.
\label{Md}
\ee
Notice that the above mass eigenvalues can always be taken 
to be real and possitive, and we  will do so hereafter.
Indeed, since right handed neutrinos have no further interactions in any other 
sector of the SM, one can always rephase right handed neutrino wave functions
to absorbe the mass phases within the $V$ matrix. In other words, such phases 
have 
not any physical meaning. 

Determination of the required mixings, $U$ and $V$, can be done by 
considering the hermitian squared matrices
\be 
\widetilde{M}_L = M_\nu M_\nu^\dagger \qquad \text{and} \qquad \widetilde{M}_R 
= M_\nu^\dagger M_\nu~.
\ee
By setting Eq.~(\ref{Md}) in above expressions, it is easy to see that
$U$ ($V$) is actually the unitary matrix that diagonalizes $\widetilde{M}_L$
($\widetilde{M}_R$), since
\be 
U^\dagger\widetilde{M}_L U = M_d^2 = diag\{m_1^2,m_2^2,m_3^2\}~,
\label{md2}
\ee
and similarly for $\widetilde{M}_R$, since 
$V^\dagger\widetilde{M}_R V = M_d^2$.

The tranformation of the left handed neutrino sector, on the 
other hand, does affect other SM sector. Specifically, in the 
neutrino mass basis, charged courrent
interactions, where the $W$ boson connects neutrinos to charged leptons, are 
now given by the coupling term
$W_\mu \bar\ell_L \gamma^\mu U_{\ell a} \nu_a$. Note that this is just as in 
the quark sector where the mixing is expressed by the CKM matrix.
In the standard parametrization the $U$ mixing matrix is given by the 
Pontecorvo-Maki-Nakagawa-Sakata 
matrix~\cite{Pontecorvo-57-58,MNS}, 
\begin{equation}
 %U = 
\left( \begin{array}{ccc} 
c_{12} c_{13} & s_{12}c_{13} & z \\
- s_{12} c_{23} - c_{12}s_{23}\bar{z}   & 
c_{12}c_{23} - s_{12}s_{23}\bar{z} & s_{23}c_{13} \\
s_{12}s_{23} - c_{12}c_{23}\bar{z} & 
- c_{12}s_{23} -c_{23}s_{12}\bar{z} & c_{23}c_{13} 
\end{array} \right),\nonumber
\end{equation}
where $z=s_{13} e^{-i\delta_{CP}}$, with $\delta_{CP}$ the
still undetermined Dirac $CP$ phase. Here, $\bar z$ stands for the 
complex conjugated, whereas $c_{ij}$ and $s_{ij}$ stand for 
$\cos \theta_{ij}$ and  $\sin \theta_{ij}$, respectively,  of the mixing angles 
with the proper identification of $\theta_{12}=\theta_\odot$, 
and $\theta_{23}=\theta_{ATM}$.

It is worth noticing that $V$ actualy contains no further physical information. 
As a matter of fact, the definition of the right handed neutrino flavor basis 
is ambiguos. Although broken, the $U(3)_N$ flavor symmetry of the 
sector allows an arbitrary redefinition of the Yukawa couplings given in 
Eq.~(\ref{yukawas}) through the trasformation 
$N_\ell\rightarrow O^N_{\ell\ell^\prime} N_{\ell^\prime}$. 
This is unlike the mass eigenstate basis, which is rather well defined. 
We make use of this fact to introduce a simple bottom up 
reconstruction of the Dirac neutrino mass matrix, based only on 
physical and measurable parameters. This truely phenomenological approximation 
would have the advantage of providing  a general and simple criteria to fix the 
weak interaction flavor basis as the one connected to the mass basis only 
through the physical rotation associated to the left handed neutrinos, given by 
the PMNS mixing matrix. 
In such a basis, we simply write
$M_\nu=U\cdot M_d$. Moreover, in this basis weak interactions are diagonal 
and the Yukawa couplings are 
expressed without lost of generality as
\be
\frac{m_\ell}{v} \bar L_\ell H \ell_R + \frac{m_a}{v}U_{\ell a} \bar L_\ell 
\widetilde{H} 
N_{a} + h.c.~,
\label{phen-yukawas}
\ee
where all unphysical parameters had been explicitely removed.
These terms provide a well defined extension to the SM that contains no 
further parameters than those already known or which can be determined in the 
future. 
The hardest to probe might be the absolute mass scale. Current bound is at 
the 2~eV range~\cite{massRange}, but if it were high enough 
we could expect to have some possitive results from tritium 
beta decay experiments~\cite{tritium,katrin}, with no positive signal from 
neutrinoless double beta decay experiments. 
Of course, the immediate form of the couplings derived from any specific 
flavor model would in general differ from  above expression, but as we have 
argued, it  can always be converted into that. 
Furthermore, above  approach has another clear 
advantage since it also allows to identify some symmetries that are hidden in 
the general expresion (\ref{yukawas}), as we discuss next.

%%%%%%%%%%%%%%%%%%%%%%%%%%%%%%%%%%%%%%%%%%%%%%%%%%%%%%%%%%%%%%%%%
\section{$\mu -\tau$ symmetry with Dirac neutrinos}
%%%%%%%%%%%%%%%%%%%%%%%%%%%%%%%%%%%%%%%%%%%%%%%%%%%%%%%%%%%%%%%%%

By using the phenomenological parameterization given in 
Eq.~(\ref{phen-yukawas}), it is strightforward to see that in the limit where
$\theta_{13}$ is null and $\theta_{23} = -\pi/4$, i.e. $\Delta\theta = 0$, 
the neutrino mass matrix exhibits a $\mu-\tau$ exchange symmetric structure,
\be 
M_\nu^0 =  \left( \begin{array}{ccc} 
m_{e1}^0 & m_{e2}^0 &0\\[1ex]
m_{\mu1}^0 & m_{\mu2}^0 & m_{\mu3}^0\\[1ex]
m_{\mu1}^0 & m_{\mu2}^0 & -m_{\mu3}^0
\end{array} \right)~. 
\label{MD}
\ee
Notice the odd behaviour of the last column. In terms of the observables, 
in our approximation the above mass terms are given as
$m_{e1}^0=m_1 c_{12}$; $m_{e2}^0=m_2 s_{12}$;
$m_{\mu 1}^0= -\frac{1}{\sqrt{2}} m_1 s_{12}$;  
$m_{\mu 2}^0= \frac{1}{\sqrt{2}} m_2 c_{12}$; and 
$m_{\mu 3}^0= - \frac{1}{\sqrt{2}}m_3$. We should stress that this $\mu-\tau$
realization is actually close related to the parametric form of the PMNS mixing 
matrix. 

Here we also note that the alternative choice for the quadrant of the  
atmospheric angle, that is taking its maximal value at $\theta_{23}= \pi/4$, 
only changes by a global sign the third row on $M_\nu^0$, which  still 
reflects a symmetric relation associated to $\mu-\tau$ exchange. This 
alternative can also be interpreted as a change in the flavor and mass state 
basis, where tau flavours and third neutrino mass eigenstate are rephased by 
$\pi$ (a simple sign change in the wave functions). This does not affect mass 
eigenvalues (they remain possitive),  neither our main conclusion about the 
size of $\mu-\tau$ breaking, as we will explicitly show later 
on for matter of completeness. In what follows we shall first concentrate in 
analizing the previously given case.

A less parametric dependant way of 
realizing the existance of $\mu-\tau$ symmetry arises when one  rather looks 
at the more 
generic form of the hermitian squared matrix, $\widetilde{M}_L$, in the 
diagonal charged lepton 
basis used for Eq.~(\ref{yukawas}). Indeed, as it can be seen from its 
definition, the 
matix form of $\widetilde M_L$ is independent of the choice made for the right 
handed neutrino basis.  Therefore, by using 
either our phenomenological basis, i.e. Eq.~(\ref{MD}), or 
directly from Eq.~(\ref{md2}), the result of calculating  the hermitian squared 
mass matrix elements, for $\theta_{13}=\Delta\theta =0$,  shows that 
$\widetilde M_L$ is symmetric and
exhibits a perfect $\mu-\tau$ exchange symmetry,
\be 
\widetilde{M}_L^0 =  \left( \begin{array}{ccc} 
\widetilde m_{ee}^0 & \widetilde m_{e\mu}^0 &\widetilde m^0_{e\mu}\\[1ex]
\widetilde m_{e\mu}^0 &\widetilde m_{\mu\mu}^0 &\widetilde m_{\mu\tau}^0\\[1ex]
\widetilde m_{e\mu}^0 &\widetilde m_{\mu\tau}^0 & \widetilde m_{\mu\mu}^0
\end{array} \right)~, 
\label{ml-mt}
\ee
where the only four relevant terms are
\be
\label{elM1}
\widetilde{m}^0_{ee} = \left\{
\ba{lc} \Delta m_{21}^2 s_{12}^2 + m_0^2 &\text{(NH)}\\[1ex]
   \Delta m_{21}^2 s_{12}^2 + |\Delta m_{31}^2| + m_0^2 &\text{(IH)}
\ea
\right. 
\ee
for normal (NH) and inverted hierarchy (IH), respectively,
 and
 \bea
 \label{elM2}
\widetilde{m}^0_{e\mu} 
&=&\frac{1}{\sqrt{2}}c_{12}s_{12}\Delta m_{21}^2 ~;
\nonumber \\
\widetilde{m}^0_{\mu\tau} &=& 
\frac{1}{2}\left( \Delta m^2_{21} c_{12}^2 \mp |\Delta m^2_{31}|\right)~;\\
\widetilde{m}^0_{\mu\mu} &=&
\frac{1}{2}\left( |\Delta m^2_{31}| +\Delta m^2_{21} c_{12}^2+ 2m_0^2\right)~;
\nonumber
\eea
where the minus (plus) sign in $\widetilde{m}^0_{\mu\tau}$ corresponds to 
NH (IH) hierarchy, and $m_0$ stands for the lighter neutrino mass. 
Notice that in this limit there is no CP violation implied, because all matrix 
elements are real in the reconstructed matrix in Eq.~(\ref{ml-mt}).
Also note that above expresions imply that, in the current  approximation, 
neutrino oscillation scales are given strightforwardly by the off diagonal 
terms of $\widetilde{M}_L^0$, with the propper identifications, such that 
$\Delta m^2_\odot = \sqrt{8} \widetilde m^0_{e\mu}/\sin 2\theta_{12}$, whereas 
$\Delta m^2_{ATM}\approx \mp 2\widetilde m^0_{\mu\tau}$. However, as this is 
just a naive approximation that neglects the contributions of $\theta_{13}$ and 
$\Delta\theta$, we need to keep in mind that it is likely to provide wrong 
predictions for the scales if the corrections from the breaking of $\mu-\tau$ 
symmetry were not negligible. We will address this issue below.

\subsection*{$\mu-\tau$ symmetry predictions}

It is not difficult to 
see that  Eq.~(\ref{ml-mt}) does correspond to  the
most general estructure of the left handed hermitian squared matrix allowed by 
$\mu-\tau$ symmetry. Indeed, in the top-down approximation where one starts by 
imposing the symmetry on the otherwise general hermitian $\widetilde M_L$, 
it is required that its elements satisfy the conditions 
$\widetilde{m}_{e\mu}=\widetilde{m}_{e\tau}$,
$\widetilde{m}_{\mu e}=\widetilde{m}_{\tau e}$,
$\widetilde{m}_{\mu\mu}=\widetilde{m}_{\tau\tau}$ and 
$\widetilde{m}_{\mu\tau}=\widetilde{m}_{\tau\mu}$. Hermiticity, on the other 
hand,  implies that all matrix elements in general obey the condition
$\widetilde{m}_{\alpha\beta}=\widetilde{m}^*_{\beta\alpha}$.
As a consequence of the last, $\widetilde{m}_{\mu\tau}$, as well as 
the diagonal components, must be real numbers. Therefore, only the off diagonal 
terms on first row and column could be complex. That is only
$\widetilde{m}_{e\mu}$ does it.  The single phase of this term, however, is non 
physical. As it can be easily checked, the last can be rephased away by a 
global redefinition of the electron neutrino, and electron wave function 
phases, which finally renders  $\widetilde{m}_{e\mu}$ to be a real number. This 
procedure shows that, indeed, no CP violation is implied in the $\mu-\tau$ 
symmetric case.

A strightforward calculation in the top-down approximation, shows that   the 
squared neutrino masses predicted from a $\mu-\tau$ symmetric $\widetilde{M}_L$ 
are given as 
\bea
{m_1}^2 &=& \frac{1}{2}\left[ \widetilde{m}^0_{ee} + \widetilde{m}^0_{\mu\mu} +
\widetilde{m}^0_{\mu\tau} - 
\frac{\sqrt{8}}{\sin 2\theta_{12}}\widetilde{m}^0_{e\mu} \right]~; \nonumber 
\\
{m_2}^2 &=& \frac{1}{2}\left[ \widetilde{m}^0_{ee} + \widetilde{m}^0_{\mu\mu} +
\widetilde{m}^0_{\mu\tau} + 
\frac{\sqrt{8}}{\sin 2\theta_{12}}\widetilde{m}^0_{e\mu} \right]~;\nonumber \\
{m_3}^2 &=& \widetilde{m}^0_{\mu\mu} - \widetilde{m}^0_{\mu\tau}  
\eea
where the $\theta_{12}$ mixing angle goes as
\be
\tan 2\theta_{12} = 
\frac{\sqrt{8}\,\widetilde{m}^0_{e\mu}}
{\widetilde{m}^0_{ee}-(\widetilde{m}^0_{\mu\mu}+\widetilde{m}^0_{\mu\tau})}~.
\ee

%%%%%%%%%%%%%%%%%%%%%%%%%%%%%%%%%%%%%%%%%%%%%%%%%%%%%%%%%%%%%%%%%
\section{$\mu -\tau$ symmetry breaking}
%%%%%%%%%%%%%%%%%%%%%%%%%%%%%%%%%%%%%%%%%%%%%%%%%%%%%%%%%%%%%%%%%

The symmetry under consideration, however, is not in any way an exact one and 
previous predictions would result to be inaqurate. 
The observed non zero values for $\theta_{13}$ and $\Delta\theta$ are a clear 
indication of that. Nevertheless, the fact that these last are actually 
small suggest that $\mu-\tau$ symmetry could still be treated as an 
approximated flavor symmetry. Exploring how good that approximation actually 
is, is the question we  address next.

In order to study the effects of the breaking of $\mu-\tau$ symmetry we will 
focus in the hermitian squared matrix $\widetilde M_L$, which, as we have 
already argued, has a general form that is independent of the chosen right 
handed neutrino basis. In 
its more general form any such a matrix can always be rewritten as
\be
\widetilde M_L = \widetilde M_L^S + \delta\widetilde M_L~,
\ee
where $\widetilde M_L^S$ is an explicitly $\mu-\tau$ exchange invariant 
matrix, and $\delta\widetilde M_L$ stands for the non invariant parts. 
In terms of its components,  the symmetric part of the hermitian 
matrix is given as
\be
\widetilde{M}_L^S =  \left( \begin{array}{ccc} 
\widetilde m_{ee} & \widetilde m_{e\mu} &\widetilde m_{e\mu}\\[1ex]
\widetilde m_{\mu e} &\widetilde m_{\mu\mu}& Re(\widetilde m_{\mu\tau})\\[1ex]
\widetilde m_{\mu e} &Re(\widetilde m_{\mu\tau}) & \widetilde m_{\mu\mu}
\end{array} \right)~.
\label{MLS}
\ee
As we have discussed in previous section above matrix can 
be made all real by a global rephasing on the electron sector. However, 
with a broken symmetry this operation can not completely remove the 
phase anymore and CP violation should arise. Indeed, the rephasing of 
electron leptons only moves the $\widetilde m_{e\mu}$ phase into 
$\delta\widetilde M_L$ matrix elements.   In order to keep our discussion 
simple and as  general as possible we assume hereafter only the natural 
conditions implied from the symmetry and hermiticity. As such, we take 
$\widetilde m_{e\mu}$ as 
the only possible complex matrix element in $\widetilde M_L^S$. Notice that 
this conditions also requires that $(\widetilde M_L^S)_{\mu\tau}$ be the real 
part of the in general  complex $\widetilde m_{\mu\tau}$. 

The nonsymmetric part, and the source of the breaking of the symmetry, is then 
expressed in general by the hermitic matrix 
\be
\delta \widetilde{M}_L =  \left( \begin{array}{ccc} 
0 &0 &\alpha\\[1ex]
0 &0 & \zeta\\[1ex]
\alpha^* &\zeta^* & \beta
\end{array} \right)~,
\label{deltaML}
\ee
where the involved symmetry breaking parameters are exactly defined as 
\bea
\alpha &=& \widetilde m_{e\tau}-\widetilde m_{e\mu}~, \nonumber \\
\beta  &=& \widetilde m_{\tau\tau}-\widetilde m_{\mu\mu}~,\\
\zeta  &=& i\, Im\,( \widetilde m_{\mu\tau})~. \nonumber
\eea
We note that hermiticity implies the existence of only one 
arbitrary phase which is contained in the $\alpha$ parameter. $\beta$ is a real 
number, 
whereas $\zeta$ is purely imaginary by definition.  From here, it is easy to 
see 
that a possible removal of the $\widetilde m_{e\mu}$ phase in 
$\widetilde M_L^S$ does only change the phase of $\alpha$, without affecting 
the other parameters. In that basis, the rephasing process 
explicitly shows that only one $CP$ phase does become physical. The relation 
of the final phase with the 
Dirac CP phase in the PMNS matrix, however, is not strightforward, as our next 
results show. Thus, we do not find any further advantage on explicitly 
using such a rephasing.

As it was  already emphasized in the discussion on previous section,
up to non physical phases, the general squared mass matrix $\widetilde M_L$  
can be reconstructed purely from neutrino 
observables using Eq.~(\ref{md2}), according to which 
$\widetilde m_{\alpha\beta} = \sum_a U_{\alpha a} U_{\beta a}^*m_a^2$. By 
comparing this reconstruction with our 
above parameterization we find that the general symmetry breaking parameters, 
written in terms of the neutrino observables without any approximation, are 
given by
\bea
\alpha &=& \mp az\mA +(as_{12}^{2}z-c_{12}s_{12}a')\mS~, \nonumber \\
\beta  &=& \pm bc_{13}^{2}\mA +[bb'+ d\cos\delta_{CP}]\mS~,   \nonumber\\
\zeta  &=& - i\, c_{12}s_{12}s_{13}\sin\delta_{CP}\mS ~, \label{sbp}
\eea
where a short hand notation has been introduced to account for the following 
combinations among the mixings,
$a=c_{13}(s_{23}-c_{23})$, $a'=c_{13}(s_{23}+c_{23})$, 
$b=c_{23}^{2}-s_{23}^{2}$, $b'=s_{13}^{2}s_{12}^{2}-c_{12}^{2}$, 
$d=4s_{12}c_{23}c_{12}s_{23}s_{13}$. In above, the sign on top (bottom) 
corresponds to NH (IH) from now on.
We stress that above expresions do cancell in the limit where
$\theta_{13} = -\pi/4$ and $\theta_{13}=0$. Furthermore, none of the breaking 
parameters do depend on the absolute scale of the neutrino mass, which seems 
remarcable.

It is also worth noticing that $\alpha$ and $\beta$ corrections are 
dominated by the atmospheric scale,  whereas $\zeta$ 
is just proportional to the solar scale.
This represents a relevant correction to the symmetric condition,
$\widetilde{m}_{e\tau}^0=\widetilde{m}_{e\mu}^0 $ as presented in  
Eq.~(\ref{elM2}).
In contrast,  the most general 
expression of such a mass term goes as $\widetilde{m}_{e\mu} = \pm 
c_{13}s_{23}z\Delta m_{ATM} + c_{13}s_{12}(c_{12}c_{23} - z s_{12}s_{23})
\Delta m_\odot$~, which  also  has the 
atmospheric scale as the leading contribution to it. This means that the 
predicted $\Delta m_{21}^2$ scale in the symmetric limit would be larger 
than the actual value of the solar scale, but not as large as the atmospheric 
scale itself (due to the $z$ factor).

In order to get a quantitative estimate of how good the  $\mu-\tau$ 
approximation is, we introduce a set of dimensionless parameters that compare 
the breaking parameters against the corresponding matrix elements, such that 
we define
\bea
\label{adim}
\hat\alpha &=&\frac{\alpha}{\widetilde m_{e\mu}} =
\frac{\sum_a U_{ea}\left[U_{\tau a}^*-U_{\mu a}^*\right]m^2_a}
{\sum_a U_{ea}U_{\mu a}^*m^2_a}~,       
\nonumber \\
\hat\beta  &=& \frac{\beta}{\widetilde m_{\mu\mu}}=
\frac{\sum_a \left[|U_{\tau a}|^2-|U_{\mu a}|^2\right]m^2_a}
{\sum_a |U_{\mu a}|^2 m^2_a} ~,\\
\hat\zeta  &=& \frac{\zeta}{Re(\widetilde m_{\mu\tau})} =
i\frac{Im \sum_a U_{\mu a}U_{\tau a}^*m^2_a}
{Re \sum_a U_{\mu a}U_{\tau a}^* m^2_a} ~. \nonumber
\eea
We should notice that this new set of parameters is invariant under  the 
rephasing of $\widetilde{m}_{e\mu}$ discussed above.

It is  strightforward to use above definitions to make an 
estimate of the leading order values of the dimensionless parameters, up to 
corrections of the order of $x={\mS}/{\mA}$. After some algebra we get  
\begin{equation}
 \hat\alpha\approx \cot\theta_{23}-1 \pm
 \left(\frac{a_1}{a_3z} +\frac{a\,a_2}{a_3^2z}\right) x +{\cal O}(x^2)
 \label{alexp1}
 \end{equation}
and 
\begin{eqnarray}
\hat\beta&\approx & \frac{\pm bc^2_{13}}{y+ a_4}-
\left[\frac{bb'+d\cos\delta_{CP}}{y+ a_4} \mp
\frac{b|a_{2}|^2}{s_{12}^2(y+ a_4)^2}\right]x \nonumber \\
&&+{\cal O}(x^2)
 \label{beesxp1}                 
\end{eqnarray}
where our short hand notation now stands for 
$y={m_{0}^{2}}/{\mA}$, 
$a_1 = as_{12}^{2}z-c_{12}s_{12}a'$, 
$a_2 = c_{13}s_{12}(c_{12}c_{23} - z s_{12}s_{23})$,  
$a_{3}=c_{13}s_{23}$
and 
\begin{equation}
 a_4 = \left\{ \ba{ll}
 s_{23}^2c_{13}^2 & \text{(NH)}\\
 1- s_{23}^2c_{13}^2 & \text{(IH)}
 \ea\right.
 \end{equation}

Considering best fit values for the observed mixings we get that 
\be
\hat\alpha\approx - 2.13 + O(x).
\ee
for any hierarchy and any value of the 
absolute neutrino mass, which represents a non small number, although $\alpha$ 
itself is always smaller than the atmospheric scale. This 
is, as a matter of fact, the largest of the corrections to the symetric mass 
matrix elements that are required to account for the observed data.
Indeed,  for $\hat\beta$ we get,  at a first order evaluation on the best fit 
values, that
\be
|\hat\beta|<\left\{\ba{ll} 
0.288 & \text{(NH)} \\
0.215 & \text{(IH)}
\ea\right.
\label{beesxp2}
\ee
It is also straightforward to show that
\begin{equation}
 \hat\zeta\approx 
 \mp i\frac{c_{12}s_{12}s_{13}\sin{\delta_{CP}}}{c_{13}^2c_{23}s_{23}}\,x+ 
{\mathcal{O}}(x^2)~.
\end{equation}
and therefore,
that $|\hat\zeta|$ has a best fit value of order $10^{-2}$ at the highest.

%%%%%%%%%%%%%%%%%%%%%%%%%%%%%%%%%
\subsection{Mixings near the symmetric limit}
%%%%%%%%%%%%%%%%%%%%%%%%%%%%%%%%%

Above results indicate that $\mu-\tau$ symmetry can only be considered as a 
good approximate flavor symmetry in a weak sense, that is  when the breaking 
parameters are compared to the heaviest neutrino mass in the spectrum, 
which is given as $m_h^2 = \mA + m_0^2 ~(+ \mS)$ for the NH (IH) case.
This claim becomes transparent if we consider the initial observation that  
$\sin\theta_{13}$ and $\sin\Delta\theta$ are, as a matter of fact, small 
numbers. 
Thus, $\alpha$ and $\beta$ 
breaking parameters can be expressed by the following
approximated first order formulae
\bea
\label{sbp2}
\alpha&\approx& 
\sqrt{2}[z(\pm\mA-s_{12}^{2}\mS)-c_{12}s_{12}s_{\Delta\theta}\mS]        
\nonumber \\
\beta &\approx& \pm 2s_{\Delta\theta}\mA +[2b's_{\Delta\theta}+gs_{13}]\mS~,
\eea
with $g=2s_{12}c_{12}(s_{\Delta\theta}^{2}-1)\cos\delta_{CP}$ and 
$s_{\Delta\theta}$ standing for 
$\sin\Delta\theta$, which for the present case is defined by the relation 
$\theta_{23}=-\pi /4+\Delta\theta$.
Above expressions do stress that indeed $|\alpha|,|\beta|, |\zeta|<<m_h^2$.
At the leading order (when $x\approx 0$) and  taking neutrino oscillation 
scales as known inputs,  we get the following predictions
\begin{equation}
 \sin\theta_{13}\approx \frac{|\alpha|}{\sqrt{2}\mA}~,
\end{equation}
and 
\begin{equation}
 \sin{\Delta\theta}\approx \frac{\beta}{2\mA}~,
\end{equation}
that should be valid for any flavor model that results consistent with neutrino 
data. 

From last expressions, the phenomenological relation given in 
Eq.(\ref{relation}) gets 
justified, since our $\mu-\tau$ parametrization now suggests that 
\begin{equation}
\sin\Delta\theta\approx \sin\theta_{13}\times \beta/\sqrt{2}|\alpha|~, 
\end{equation}
which 
is an expression given just in terms of the mass matrix elements of 
$\widetilde{M}_L$.

Including solar scale contribution in our calculations provides  somewhat more 
complicated expresions for the small mixings that cannot be easily resolved 
analiticaly in terms only of the mass matrix elements. However, one can 
read Eq.~(\ref{sbp2}) as a constraint on the small neutrino mixings, when all 
other neutrino oscillation parameters are taken as known, within experimental 
uncertainties. Following this line of thought, and after some lenghtly algebra 
we can express the relation among the deviation of the atmospheric mixing from 
its maximal value in terms of the predicted value for $\theta_{13}$, as
\be
\sin\Delta\theta \approx 
\sin\theta_{13}\frac{g_{\delta}(\sqrt{2}Bg_{s}+As_{12}c_{12}\mS)}
                    {2Ag_{c}+\sqrt{2}Bs_{12}c_{12}\mS}~, 
\label{phenrel}
\ee
where we have defined 
$A = \mathrm{Re}(\alpha)+\mathrm{Im}(\alpha)$,  
$B =\beta +|\zeta|$, and also 
$g_{s}=\pm\mA-\mS s_{12}^{2}$,
$g_{c}=\pm\mA-\mS c_{12}^{2}$, and 
$g_{\delta}= \cos\delta_{CP}-\sin\delta_{CP}$.
In the same footing, the predicted value for $\sin\theta_{13}$ mixing is 
given as
\be
\sin\theta_{13}\approx \frac{2Ag_{c}+\sqrt{2}Bc_{12}s_{12}\mS}                 
    {2\sqrt{2}g_{\delta}[g_{c}g_{s}-(s^2_{12}c^2_{12}\mS)^2]}.
\ee

%%%%%%%%%%%%%%%%%%%%%%%%%%%%%%%%%%%%%%%%%%%%%%%%%%%%

\section{$\mu-\tau$ anti-symmetry case.}

Let us next add some comments regarding the case where $\theta_{23}$ is chosen 
to lay in the first quadrant, such that its maximal value corresponds to 
$\theta_{23} = \pi/4$. As we have  stated before this alternative should
corresponds to the one we have already discussed, up to wave function phase 
redefinitions in both flavor and mass neutrino basis. Therefore 
one would not expect any fundamental conclusion to change. Nevertheless, as the 
given choice effectively affects the way $\mu-\tau$ symmetry realizes, and, 
hence, general formulae may also change accordingly,  we believe 
that considering in some detail the changes introduced in the analysis for this 
choice can be of interest for model building. 

First of all, for a strictly positive value of $\theta_{23}=\pi/4$,
joint to a zero value for $\theta_{13}$, our phenomenological reconstruction of 
the mass matrix, in the diagonal charged lepton basis, now leads to
\be
\label{MDa}
M_{A\nu}^{0}=\left(\begin{array}{ccc}
m_{e1}^{0} & m_{e2}^{0} & 0 \\
m_{\mu 1}^{0} & m_{\mu 2}^{0} & m_{\mu 3}^{0} \\
-m_{\mu 1}^{0} & -m_{\mu 2}^{0} & m_{\mu 3}^{0} \\
\end{array}\right)~.
\ee
Comparing with Eq.~(\ref{MD}) one can notice that only the third row of the 
above mass matrix has changed  by a global sign. However, this does 
indeed change the way $\mu-\tau$ manifests itself in the hermitian squared 
mass matrix, as usually defined by 
$\Mt_{AL}^{0}=M_{A\nu}^{0}M_{A\nu}^{0\dagger}$, which now becomes
\be
\label{MDasq}
\Mt_{AL}^{0}=\left(\begin{array}{ccc}
\mt_{ee}^{0} & \mt_{e\mu}^{0} & -\mt_{e\mu}^{0} \\
\mt_{\mu e}^{0} & \mt_{\mu\mu}^{0} & \mt_{\mu\tau}^{0} \\
-\mt_{e\mu}^{0} & \mt_{\mu\tau}^{0} & \mt_{\mu\mu}^{0} \\
\end{array}\right),
\ee
where its entries are given by equations (\ref{elM1}) and (\ref{elM2}) just as 
in the symmetric case. Above matrix now exhibits a $\mu-\tau$ antisymmetry, 
where $\Mt_{AL}^{0}$ remains invariant under the exchange 
$\nu_\mu\leftrightarrow -\nu_\tau$.

It is important to point out two things in here. First, the squared mass matrix 
is also 
hermitian in the $\mu-\tau$ anti-symmetric case, and second, 
that $\Mt_{AL}^{0}$ can also be constructed by a topdown method akin to the 
$\mu-\tau$ symmetric case using the PMNS 
mixing matrix elements by considering $\theta_{23}=\pi/4$ instead of 
$\theta_{23}=-\pi/4$.
Following our own previous steps,  in the most 
general case $\mu-\tau$ anti-symmetry breaking can be explicitly parameterized 
as  
\be
\Mt_{\nu}=\Mt_{L}^{A}+\delta\Mt_{A}.
\ee
where the $\Mt_{L}^{A}$ is the $\mu-\tau$ antisymmetric part generaly written as
\be
\label{anSymM}
\Mt_{L}^{A}=\left(\begin{array}{ccc}
\mt_{ee} & \mt_{e\mu} & -\mt_{e\mu} \\
\mt_{\mu e} & \mt_{\mu\mu} & Re(\mt_{\mu\tau}) \\
-\mt_{e\mu} & Re(\mt_{\mu\tau}) & \mt_{\mu\mu} \\            
\end{array}\right).
\ee
As for the anti-symmetry breaking matrix, $\delta\Mt_{A}$, it is clear that the 
only entry 
which is diferent from the symmetric case, is the one associated to  
$\mt_{e\tau}$, and thus it is now written as
\be
\label{anSymDM}
\delta\Mt_{A}=\left(\begin{array}{ccc}
0 & 0 & \alpha_{A} \\
0 & 0 & \zeta \\
\alpha_{A}^{*} & \zeta^{*} & \beta \\
\end{array}\right).
\ee
for which we now define
\be
\label{alphaA1}
\alpha_{A}=\mt_{e\tau}+\mt_{e\mu}~,
\ee
whereas $\beta$ and $\zeta$ are given as before. In terms of the neutrino 
oscillation parameters we now get
\be
\alpha_{A}=\pm a'z\mA-(ac_{12}s_{12}+a's_{12}^{2}z)\mS.
\ee
Thus, the corresponding dimensionless parameter,  defined as earlier by 
$\hat\alpha_{A} ={\alpha_{A}}/{\widetilde m_{e\mu}}$, is given, upto first 
order in the neutrino oscillation scales ratio, as 
\begin{eqnarray}
\hat\alpha_{A}&\approx&1+\cot\theta_{23}-
 \left(\frac{a_1'}{a_3z} +\frac{a\,a_2}{a_3^2z}\right) x +{\cal O}(x^2)
\nonumber \\
&\approx& 2.13 +{\cal O}(x)
 \end{eqnarray}
where
$a_1' = a's_{12}^{2}z+a c_{12}s_{12}$. Last row gives the leading order 
contribution with central values in neutrino mixings, which is, up to a 
sign, as large as the one obtained in the symmetric realization of 
$\mu-\tau$ symmetry and,  as before, $\alpha_A$ remains 
smaller than the largest neutrino mass. 

By considering the near to anti-symmetric case, where we parameterize 
the atmospheric mixing as 
$\theta_{23}=\pi/4 +\Delta\theta$ for $|\Delta\theta|\ll 1$, we get the 
following expresion for the predicted relationship among  
$\theta_{13}$ and $\Delta\theta$ mixings, 
\be
\sin\Delta\theta\approx\sin\theta_{13}
\frac{\sqrt{2}Bg_{\delta}g_{s}-2A'g'_{\delta}s_{12}c_{12}\mS}
   {\sqrt{2}Bc_{12}s_{12}\mS-2A'g_{c}},
\ee
where $A'= \mathrm{Re}(\alpha_{A})+\mathrm{Im}(\alpha_{A})$ and 
$g'_{\delta} = \cos{\delta_{CP}}+\sin{\delta_{CP}}$. Notice that  this expression
mimics the corresponding one that  we obtained in the previous case 
[Eq.~(\ref{phenrel})]

%%%%%%%%%%%%%%%%%
\subsection{Reconstructed mass matrix elements}
%%%%%%%%%%%%%%%%%%%%%%%%%%

By using the general parameterization we have given for the hermitian squared 
mass matrix, it is easy to get a numerical idea of the order of magnitude of 
te off-diagonal matrix elements. This is because it turns out that they do not 
depend (as we already enphasized along our previous discussions) on the absolute 
mass scale of the neutrino. 
For this we can just set in the best fit  values of 
the oscillation neutrino parameters, assuming no CP violation. 
Thus,  
the following are the so numerically reconstructed off-diagonal mass matrix 
elements, in the most general case, for NH (IH),
\bea
\widetilde{m}_{\mu e} &\approx& 2.65\times 10^{-4} (2.16\times 10^{-4}) 
\mathrm{eV}^{2}, \nonumber\\
\widetilde{m}_{\mu\tau} &\approx& 1.19\times 10^{-3} (1.24\times 10^{-3}) \mathrm{eV}^{2}, \\
\widetilde{m}_{\tau e} &\approx& 2.46\times 10^{-4} (2.96\times 10^{-4}) \mathrm{eV}^{2}. \nonumber
\eea

On the other hand, the reconstruction of the  
diagonal terms in the CP conserving case gives the 
following expressions, 
{\small
\bea
\widetilde{m}_{ee} &=& 
m_{0}^{2}+\mA\left[\left(\frac{1}{2}\mp\frac{1}{2}\right)\pm 
                   s_{13}^{2}\right]+s_{12}c_{13}^{2}\mS, \nonumber\\
\widetilde{m}_{\mu\mu} &=& m_{0}^{2}+\mA\left[\left(\frac{1}{2}\mp\frac{1}{2}\right)
                       \pm s_{23}^{2}c_{13}^{2}\right]+h\mS,\nonumber\\
\widetilde{m}_{\tau\tau} &=& m_{0}^{2}+\mA\left[\left(\frac{1}{2}\mp\frac{1}{2}\right)
                       \pm c_{23}^{2}c_{13}^{2}\right]+h'\mS, \nonumber
\eea}
where, to simplify, we have  used
$h = 
c_{12}^{2}c_{23}^{2}+s_{12}^{2}s_{23}^{2}s_{13}^{2}-2c_{12}c_{23}s_{12}s_{23}s_{
13}
      \cos\delta_{CP},$ %\nonumber\\
and 
$h' = 
c_{12}^{2}s_{23}^{2}+s_{12}^{2}c_{23}^{2}s_{13}^{2}+2c_{12}c_{23}
s_{12}s_{23}s_{13} \cos\delta_{CP}$~,
and where the top(bottom) sign corresponds
to normal (inverted) hierachy and $m_{0}$ stands for the lightest neutrino 
mass, as usual.

\section{Higgs boson decay to neutrinos.}

From the Yukawa interaction in the flavor basis, as expressed by equation 
(\ref{phen-yukawas}), it is 
straightforward to compute 
the invisible Higgs boson decay width, 
$\Gamma_{a}^{\ell}(H\rightarrow\nu\overline{\nu})$.
Also, considering the current observed value of the total Higgs' decay width 
$\Gamma_{H}<1.7$ GeV~\cite{PDG2016}, the branching ratio for the invisible decay
$B(H\rightarrow\nu\overline{\nu})$, to first order,
can be estimated within the current framework.
For the decay width, considering that $m_{a}\ll m_{H}$, we get 
\be
\label{DecWidNeu}
\Gamma_{a}^{\ell}\approx
\frac{m_{a}^2}{2\langle v\rangle^2}\frac{m_{H}}{8\pi}|U_{\ell a}|^2 ~.
\ee
The branching ratio is then given by the equation
\be
\label{branchneut}
B(H\rightarrow\nu\overline{\nu})=\frac{\sum_{a,\ell}\Gamma_{a}^{\ell}}{\Gamma_{H}}~.
\ee

Furthermore, regarding the Higgs mass at a value of $m_{H}=125$ GeV, its vev 
$\langle v\rangle\approx 246$ eV~\cite{HiggsMass} and taking the neutrino mass spectrum at 
a high limit of $m_{a}\approx 2$ eV in a degenerated hierarchy as a means of estimating an 
upper limit value, the resulting invisible branching ratio comes out to be 
\be
B(H\rightarrow\nu\overline{\nu})\approx 5.5\times 10^{-13}, 
\ee
at its larger possible value in the theory.

The upper bound experimentaly set for the branching fraction of the invisible 
Higgs decay  is currently at 0.28~\cite{ATLASco}, which places our current 
estimation at a much lower value.

%%%%%%%%%%%%%%%%%%%%%%%%%%%%%%%%%%%%%%%%%%%%%%%%%%%%%%%%%%%%%%%%%
\section{Conclusions}

As our present analysis for Dirac neutrinos has shown, $\mu$-$\tau$ symmetry 
arises as a slightly broken symmetry that makes itself evident in the 
diagonal charged lepton mass basis. This symmetry is actually already encoded 
by the observed mixings in the PMNS mixing matrix. 
An appropriate selection of the right 
handed neutrino basis then allows to remove all non physical parameters in the 
Yukawa sector. In such a parameterization, only the absolute neutrino 
mass scale and the CP phase remain unknown. 

Furthermore,  $\mu$-$\tau$ breaking becomes  easier to study when  using the 
hermitian squared neutrino mass matrix ($\widetilde{M}$). Such a matrix can 
either be symmetric or antysymmetric under the exchange of $\mu$ and 
$\tau$ labels. Both the realizations, however, as we have already argue,  are 
actually connected with a simple rephasing on $\tau$ neutrino flavor and third 
neutrino mass eigenstates and, therefore, main conlcusions regarding symmetry 
breaking are alike.
$\widetilde{M}$ allows for a natural and easy 
parameterization of the breaking of the symmetry that requires only three free 
parameters ($\alpha$, $\beta$ 
and $\zeta$), with only one of them being a complex number ($\alpha$). 
Observed atmospheric and reactor mixings indicate that the 
symmetry breaking parameters are at most below the range of the 
atmospheric squared mass difference, without any dependance on the absolute 
scale of the neutrino mass and regardless of the actual value of the CP phase.

The largest of the symmetry breaking corrections, in any realization of the 
symmetry, corresponds to the $\widetilde{m}_{e\tau}$ mass term, paremeterized 
by $\alpha$, which in any case is always smaller than the heavier mass scale 
in the neutrino sector. 
This could indicate the presence of an unknown (perhaps broken) symmetry 
in the $e-\tau$ coupling sector.
It is also interesting to note that a perturbative behavior 
in the symmetry 
breaking sector only happens when the neutrino mass spectrum becomes almost 
degenerate and, thus, when the absolute mass scale turns out to be larger than 
the atmospheric scale.
Indeed, upon comparison of the matrix elements in $\delta\Mt_{\alpha\beta}$ 
with the neutrino mass scale, it is only possible for the almost degenerate 
case  to truely fulfill  the relation $\delta\Mt_{\alpha\beta}\ll m_{0h}^{2}$ 
with $m_{0h}^{2}$ being the 
squared minimum value of the neutrino mass scale. 
Therefore,  the only general conclusion one can draw is that $\mu-\tau$ 
symmetry is indeed a good approximated symmetry only in a weak sense. 
Nevertheless, it is still interesting to point out the approximated relation 
that arises for the observed mixings with the atmospheric scale, according to 
which $2\mA \approx 
\sqrt{2}|\alpha|/\sin\theta_{13}\approx 
|\beta|/ \sin{\Delta\theta}$. 
We believe this observations may be valuable for model building in the Dirac 
neutrino framework. For further study, an analysis of the $\mu$-$\tau$ symmetry 
breaking can be performed using partubation theory considering the results 
mentioned earlier.

Another noteworhty result, is the fact that for any realization of the 
symmetry (that is for a  symmetric or  
anti-symmetric $\widetilde{M}$), we always have that 
$\hat\alpha,\hat\alpha_{A}>>\hat\beta,\hat\zeta$, 
which might also indicate an underlying 
flavor symmetry involving the $e-\mu$ coupling in the theory.
Thus, in general, even though $\mu-\tau$ symmetry seems to be an underlying 
symmetry in the neutrino sector a complete understanding of neutrino 
oscillation parameters stills seems to need additional (extended ) flavor 
symmetries.

Finally, the estimation or the invisible Higgs decay to neutrinos could grant a good test 
for the nature of neutrinos as well as their mass spectrum, mass hierarchy and CP violation
phase, provided that such quantity could be measured accurately. 
Unfortunately, however, it still lays far below the current experimental 
sensitivity.

%%%%%%%%%%%%%%%%%%%%%%%%%%%%%%%%%%%%%%%%%%%%%%%%%%%%%%%%%%%%%%%%%%

\section*{Acknowledgments}
This work was partially supported
by CONACyT, M\'exico, under Grant No. 237004.

%%%%%%%%%%%%%%%%%%%%%%%%%%%%%%%%%%%%%%%%%%%%%%%%%%%%%%%%%%%%%%%%%%%%%

\end{document}